\documentclass[twocolumn,aps,superscriptaddress,prL,floatfix,showpacs,longbibliography]{revtex4-2}

\makeatletter
\newcommand*{\rom}[1]{\expandafter\@slowromancap\romannumeral #1@} %roman number
\makeatother
\usepackage{gensymb} % \celsius
\usepackage{graphicx}% Include figure files
\usepackage{times}
\usepackage{dcolumn}
\usepackage{hyperref}
\usepackage{color}
\usepackage{float}
\usepackage{amsmath, amssymb}
\usepackage{courier}
\usepackage{xr-hyper}
\usepackage{ulem}
\usepackage{textcomp,mathcomp}

\newcommand\LNO[0]{La$_3$Ni$_2$O$_7$}
\newcommand\LPNO[0]{La$_{1.9}$Pr$_{1.1}$Ni$_2$O$_{6.97}$}
\newcommand\LNOO[0]{La$_3$Ni$_2$O$_{6.63}$}
\newcommand\musr[0]{$\mu^+$SR}

\hypersetup{colorlinks = true, linkcolor = blue, anchorcolor =blue, citecolor = blue, filecolor = blue, urlcolor = blue,
	pdfauthor=author}
\begingroup
\begin{document}
\title{Evolution of magnetism in Ruddlesden-Popper bilayer nickelate revealed by muon spin relaxation}
	
\author{Kaiwen Chen}
\altaffiliation{These authors contributed equally to this work.}
\affiliation{State Key Laboratory of Surface Physics, Department of Physics, Fudan University, Shanghai 200438, China}
\author{Xiangqi Liu}
\altaffiliation{These authors contributed equally to this work.}
\affiliation{School of Physical Science and Technology, ShanghaiTech University, Shanghai 201210, China}
\author{Ying Wang}
\altaffiliation{These authors contributed equally to this work.}
\affiliation{State Key Laboratory of Surface Physics, Department of Physics, Fudan University, Shanghai 200438, China}
\author{Ziyi Zhu}
\affiliation{School of Materials,Zhengzhou University of Aeronautics, Zhengzhou 450046,China}
\author{Jiachen Jiao}
\affiliation{State Key Laboratory of Surface Physics, Department of Physics, Fudan University, Shanghai 200438, China}
\author{Chengyu Jiang}
\affiliation{State Key Laboratory of Surface Physics, Department of Physics, Fudan University, Shanghai 200438, China}
\author{Yanfeng Guo}
\affiliation{School of Physical Science and Technology, ShanghaiTech University, Shanghai 201210, China}
\affiliation{ShanghaiTech Laboratory for Topological Physics, Shanghai 201210, China}
\author{Lei Shu}
\email{leishu@fudan.edu.cn}
\affiliation{State Key Laboratory of Surface Physics, Department of Physics, Fudan University, Shanghai 200438, China}
\affiliation{Shanghai Research Center for Quantum Sciences, Shanghai 201315, People`s Republic of China}
	%\renewcommand{\thefootnote}{\fnsymbol{footnote}}
	%\footnotetext[1]{These authors contributed equally to this  work.}
	
\begin{abstract} 
Here we report the positive muon spin relaxation study on Pr-doped \LPNO\ and oxygen-deficient \LNOO\ polycrystalline  under ambient pressure. Zero-field \musr\ experiments reveal the existence of bulk long-range magnetic order in \LPNO\ with $T_{N}=161\ \rm{K}$, while \LNOO\ exhibits a short-range magnetic ground state with $T_N=30\ \rm{K}$. The magnetic transition width of \LPNO\ revealed by weak-transverse-field \musr\ is narrower compared to La$_3$Ni$_2$O$_{6.92}$. Our \musr\ experiment results provide a comprehensive view on the correlation between magnetism and structure perfection in Ruddlesden-Popper bilayer nickelates under ambient pressure.
\end{abstract}
	 
\maketitle
\date{}

\textit{Introduction.}--- The recently discovered superconductivity with $T_c\approx80\ \rm{K}$ in pressurized Ruddlesden-Popper (RP) bilayer \LNO\ provides an exciting new platform for investigating pairing mechanism of unconventional superconductivity~\cite{Sun2023}. Subsequently studies further revealed the superconductivity under pressure in trilayer RP phase La$_4$Ni$_3$O$_{10}$ and Pr$_4$Ni$_3$O$_{10-\delta}$ with $T_c\sim30$ K~\cite{Zhaojun2024,QingLi2024,Sakakibara2024,MingxingZhang2024,Li2024,Huang2024,Wang_Review_2024}, paving the way towards more potential superconducting candidates.	Various measurements including resonant inelastic x-ray scattering (RIXS)~\cite{RIXS}, muon spin relaxation (\musr)~\cite{Zurab,Chen2024} and nuclear magnetic resonance (NMR)~\cite{NMR2024}, have revealed the presence of spin density waves (SDW) below $\sim$ 150~K in La$_3$Ni$_2$O$_{7-\delta}$ at ambient pressure, while no long-range magnetic order was observed in powder neutron scattering measurement down to 10~K~\cite{NPD2024}. Meanwhile, pressurized \musr\ experiment revealed an increase in the SDW transition temperature, in contrast to the earlier transport measurements~\cite{Sun2023,Wang2024,Yuan2023,Jing2024}. Though the nature of SDW is still controversial, most theoretical models suggest that pairing is driven by magnetic interactions~\cite{Liu_PRL2023,Qin2023,Sakakibara2023,Shen2023,Xue2024,Yang2023_theory,YYF2023,Zhang2023_theory,Luo2024,Fan2024,Heier2024,Jiang2024,Liu2023_theory,Lechermann2024}.

In addition, the RP bilayer La$_3$Ni$_2$O$_{7-\delta}$ reported so far generally contains mixed phases and randomly distributed oxygen defects~\cite{MEP,Wang2024}, which causes sample-dependent issues. Zero resistance is confirmed in both single crystals~\cite{Yuan2023,Hou2023,Jing2024,Zhou2024} and polycrystalline samples~\cite{Wang2023,Wang2024}, while superconducting volume fraction varies from 1\% to 48\%\ among different samples, indicating a decisive role of crystal imperfections in pressurized superconductivity. Substitute La with Pr can effectively improve the purity of bilayer structure, resulting higher $T_c$ and superconducting volume fraction~\cite{Wang2024}. Similar to the early-stage studies of La$_3$Ni$_2$O$_{7-\delta}$, no evidence of magnetic ground state can be revealed from magnetization or transport measurements of La$_2$PrNi$_2$O$_7$. On the other hand, no sign of superconductivity was observed in oxygen-deficient sample La$_3$Ni$_2$O$_{6.45}$ up to 41 GPa~\cite{Cava}.  Meanwhile, a broad magnetic transition near 50 K was observed under ambient pressure, whose nature is still unclear. Therefore, La$_2$PrNi$_2$O$_7$ and La$_3$Ni$_2$O$_{6.45}$  provide an ideal comparison to study the evolution of magnetism with crystal imperfectness in RP bilayer nickelates. The former is on the side of a perfect bilayer structure while the latter is on the side of extreme oxygen defects.
	
\musr\ is a powerful technique for studying microscopic magnetism. The implanted muons act as local magnetic probes and can detect small magnetic fields in the sample~\cite{Adrian2022,Blundell1999,Yaouanc2011MuonSR,Hayano1979}. Here we report the \musr\ measurements on Pr-doped \LPNO\ and oxygen-deficient \LNOO\ polycrystalline under ambient pressure. The ZF-\musr\ measurements reveal the long-range magnetic order in \LPNO\ under 161 K, while a short-range magnetic order is observed in \LNOO\ below 30 K. Combined with previous \musr\ result on La$_3$Ni$_2$O$_{6.92}$~\cite{Chen2024}, we found that \LPNO\ exhibits higher $T_N$ and narrower magnetic transition width. The exponent $\beta=0.33$ is close to value expected in the 3D Ising model. On the other hand, the long-range order is completely disrupted in the oxygen-deficient \LNOO. The properties mentioned above can be explained from the perspective of oxygen vacancies. Our \musr\ results provide a comprehensive diagram of ambient-pressure magnetism in RP bilayer nickelates.
	
\textit{Experimental details.}--- Polycrystalline samples of \LPNO\ and La$_3$Ni$_2$O$_7$ were synthesized through the solid-state reaction~\cite{Chandrasekharan2022,Wang2024}. High-purity La$_2$O$_3$ (99.999\%, Aladdin), NiO (99.99\%, Macklin), and Pr$_6$O$_{11}$ (99.99\%, Aladdin) were mixed with a molar ratio of 6:12:1 and 3:4, respectively. The ground mixtures were sintered at 1100 $\tccelsius$ in the air for 80 h. The reactants were reground and annealing at 900 $\tccelsius$ for 3 times to get completely reacted and homogeneous polycrystalline samples. After that, the sample was sintered in air at 1050 $\tccelsius$ for 20 hours. \LNOO\ was prepared by the reduction of the La$_3$Ni$_2$O$_7$ powder in flowing 10\% H$_2$ in Ar at 400 $\tccelsius$ for 5 h in a tube furnace.
	 
Powder x-ray diffraction (PXRD) patterns were measured by a Bruker D8 advanced x-ray diffraction spectrometer ($\lambda$ = 1.5418 \r{A}) at room temperature. The XRD Rietveld refinement was conducted with \texttt{FULLPROF} software~\cite{FULLPROF}. The analysis of chemical composition and microstructure used a Phenom Pro scanning electron microscope equipped with an energy-dispersive x-ray (EDX) spectrometer. Thermogravimetric analysis (TGA) measurement was used to determine the oxygen content by using a 10\% H$_2$/Ar gas flow of 50 mL/min, with a 5 $\tccelsius$/min rate up to 950 $\tccelsius$ for \LNOO\ and 7.5 $\tccelsius$/min rate up to 1050 $\tccelsius$ for \LPNO.
The magnetization measurement was conducted using a superconducting quantum interference device magnetometer (Quantum Design magnetic property measurement system). The temperature-dependent susceptibilities between 2 and 300 K were measured under a magnetic field of 0.1 T in both zero-field cooled (ZFC) and field-cooled (FC) procedures. Temperature dependence of resistivity, denoted as $\rho(T)$, was  measured using standard four-probe method with a physical property measurement system (PPMS). Powder samples were pressed and cut into rectangle. \LPNO\ sample was sintered at 1050 $\tccelsius$ for 7 h before resistivity measurement. The polycrystalline \LNOO\ is unstable under heating and the resistivity was measured without sintering. Conductive silver adhesives were applied to glue four golden wires on the surface of the samples. The resistivities were measured between 2 K and 300 K during heating procedure.  
	 
The ZF, LF, and wTF-\musr\ experiments were carried out on the LAMPF spectrometer of M20 beamline, TRIUMF. Approximately 300 mg of polycrystalline sample was pressed into pellet with a diameter of 1.2 cm. The sample was wrapped with silver tape and mounted on a hollow square copper frame, which made it free from background contributions. Both ZF and wTF-\musr\ experiments were carried out between 2.4 K and 300 K. The longitudinal field used in LF-\musr\ experiments is up to 0.4 T. The \musr\ data was analyzed with the \texttt{MUSRFIT} software package~\cite{MUSRFIT}. The fitting is limited to data after 0.03~\textmu s.
	 
\begin{figure}[h]
\begin{center}
\includegraphics[width=0.5\textwidth]{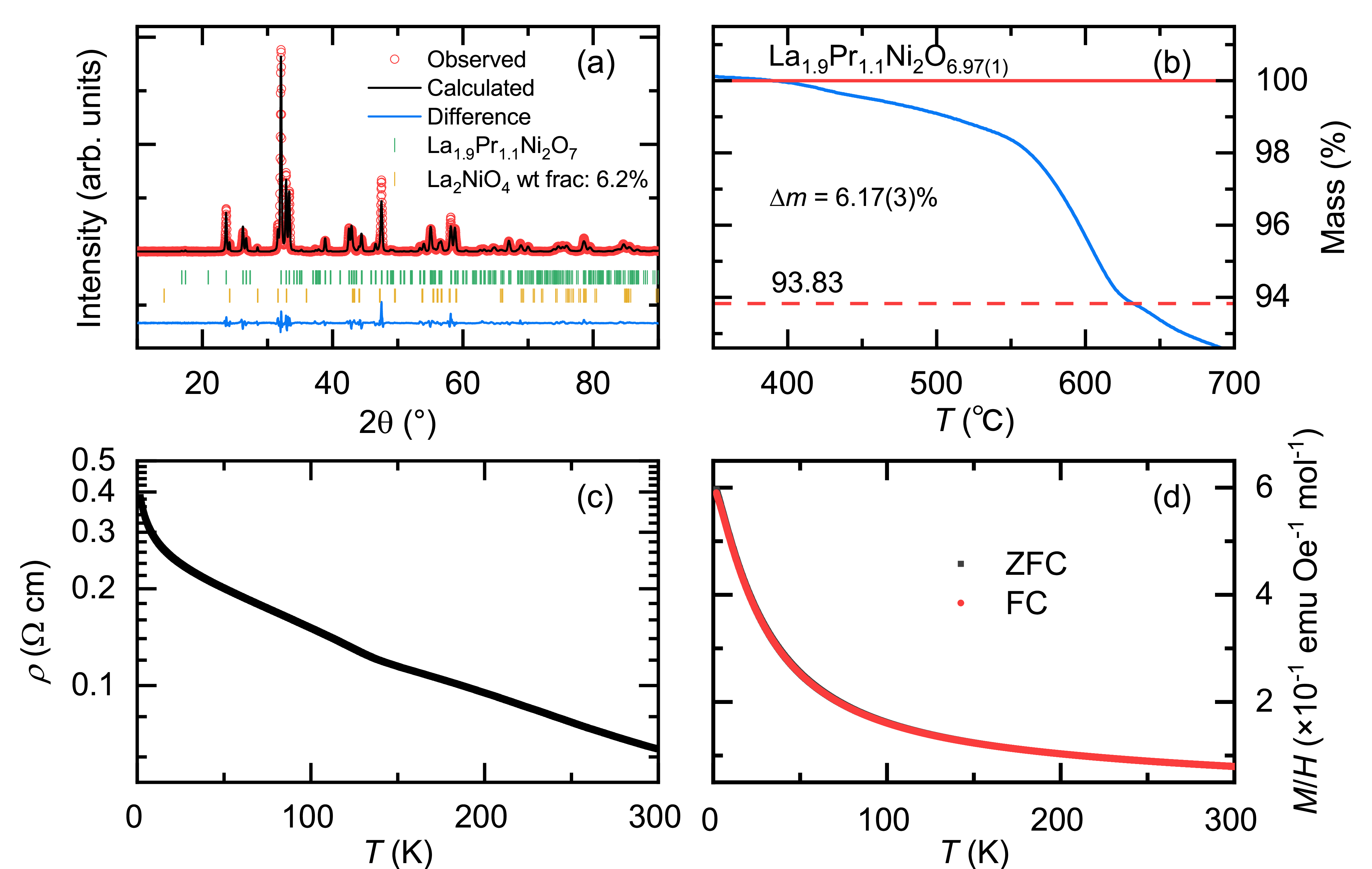}
\caption{\label{fig:Fig1}(a) Rietveld refinement of powder x-ray diffraction data of \LPNO. The refinement reveals the existence of La$_2$NiO$_4$ phase with weight fraction of 6.2(1)\%. (b) Thermogravimetric curves for \LPNO\ in 10\% H$_2$/Ar from 350 °C to 700 °C. (c) temperature dependence of the resistivity $\rho(T)$. (d) Temperature dependence of magnetic susceptibility $\chi(T)$ with $\mu_{\rm{0}}H=0.1$ T. There is no sign of spin freezing or magnetic transition down to 2 K.}
\end{center}
\end{figure}

\textit{Physical Properties}---Fig.~\ref{fig:Fig1}(a) shows the x-ray diffraction pattern of polycrystalline \LPNO. The Rietveld refinement confirms an orthorhombic structure (space group \textit{Amam}, no. 63) with $\textit{a}=5.37608(19)$ $\rm{\r{A}}, \textit{b}=5.4518(2)\ \rm{\r{A}}$, $\textit{c}=20.3847(9)\ \rm{\r{A}}$. The refinement reveals the existence of La$_2$NiO$_4$ with weight fraction of 6.2(1)\%. As shown in Fig.~\ref{fig:Fig1}(b), the oxygen stoichiometry of the initial phase is determined by calculating the weight loss between the phases in TGA measurement, which results in {La$_{1.9}$Pr$_{1.1}$Ni$_2$O$_{6.97(1)}$}. The edge of the transition is determined by the first derivative of the mass change~\cite{Supmat}.
The resultant products  $\rm{La_2O_3}$, $\rm{Pr_2O_3}$ and $\rm{Ni}$ are confirmed by powder XRD measurement. Temperature dependence of resistivity $\rho(T)$ of polycrystalline \LPNO\ is plotted in Fig.~\ref{fig:Fig1}(c).  No sign of transition can be observed on $\rho(T)$. Fig.\ref{fig:Fig1}(d) displays the temperature dependence of dc magnetic susceptibility $\chi(T)$, which was measured under the magnetic field of $\mu_0H=0.1$ T with both zero-field cooling and field cooling setup. There is no divergence between ZFC and FC curves and no sign of any magnetic phase transition or spin freezing behavior down to 2 K. The field dependence of magnetization curve further exhibits the minimal impact of La$_2$NiO$_4$~\cite{Supmat}. 
	 
\begin{figure}[h]
\begin{center}
\includegraphics[width=0.5\textwidth]{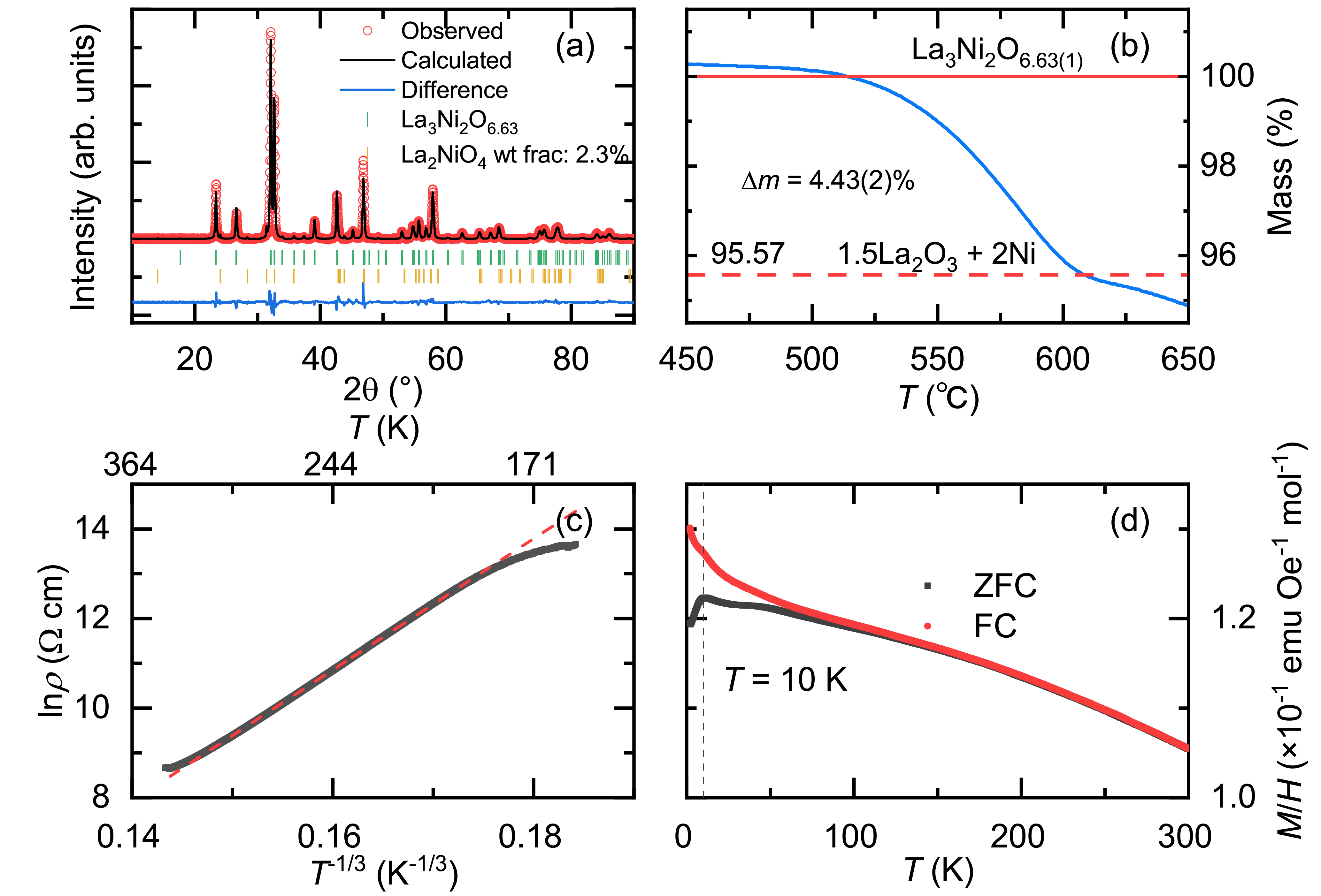}
\caption{\label{fig:Fig2}(a) Rietveld refinement of powder x-ray diffraction data of \LNOO. The refinement reveals the existence of La$_2$NiO$_4$ phase with weight fraction of 2.3(1)\%. (b) Thermogravimetric curves for \LNOO\ in 10\% H$_2$/Ar from 450 °C to 650 °C. (c) temperature dependence of the resistivity $\rho(T)$. The dashed line exhibits the fitting with Eq.~(\ref{eq:1}) between 185 K and 315~K. (d) Temperature dependence of magnetic susceptibility $\chi(T)$ with $\mu_{\rm{0}}H=0.1$ T. The magnetization curves split below 100 K between FC and ZFC setup. }
\end{center}
\end{figure}

The results of \LNOO\ are summarized in Fig.~\ref{fig:Fig2}. Fig.~\ref{fig:Fig2}(a) shows the x-ray diffraction pattern of polycrystalline \LNOO. The Rietveld refinement indicates the tetragonal structure (space group \textit{I4/mmm}, no. 139) with $a=b=3.8881(2)\ \rm{\r{A}}$, $c=20.1163(12)\ \rm{\r{A}}$. The refinement also reveals the existence of La$_2$NiO$_4$ with weight fraction of 2.3(1)\%. The oxygen stoichiometry of \LNOO\ is determined by the TGA measurement. Temperature dependence of resistivity $\rho(T)$ of polycrystalline \LNOO\ is plotted in Fig.~\ref{fig:Fig2}(c). 
Below 165 K, the resistivity exceeded the measuring range of our equipment. $\rho(T)$ between 315 K and 185 K can be fitted to Mott's law for variable range hopping (VRH) for a 2D case~\cite{Mott1968}: 
\begin{eqnarray}
\label{eq:1}
\rho=\rho_0\exp\left\{\left(\frac{T_0}{T}\right)^{\frac{1}{3}}\right\},
\end{eqnarray}
 where $T_0=3.1\times10^{6}$ K is a measure of localization and $\rho_0=2.9\times10^{5}\ \Omega\cdot\rm{cm}$  is a resistivity factor.
 The fit of $\rho(T)$ to Eq.~(\ref{eq:1}) could be related to the Anderson localization due to oxygen vacancy disorder in the sample. Fig.\ref{fig:Fig2}(d) displays the temperature dependence of dc magnetic susceptibility $\chi(T)$ measured under the magnetic field of $\mu_0H=0.1$ T. The FC and ZFC magnetization curves split below 100 K and the ZFC curve exhibits a transition at 10 K. The field dependence of magnetization curve at 300 K reveal an ferromagnetic component~\cite{Supmat}. Combined with the wTF-\musr\ experiments discussed later, we conclude that this ferromagnetic signal is not due to impurities, but rather may be related to the localized moments in the vicinity of oxygen-deficiencies.
	 
\begin{figure}[h]
	\begin{center}
		\includegraphics[width=0.5\textwidth]{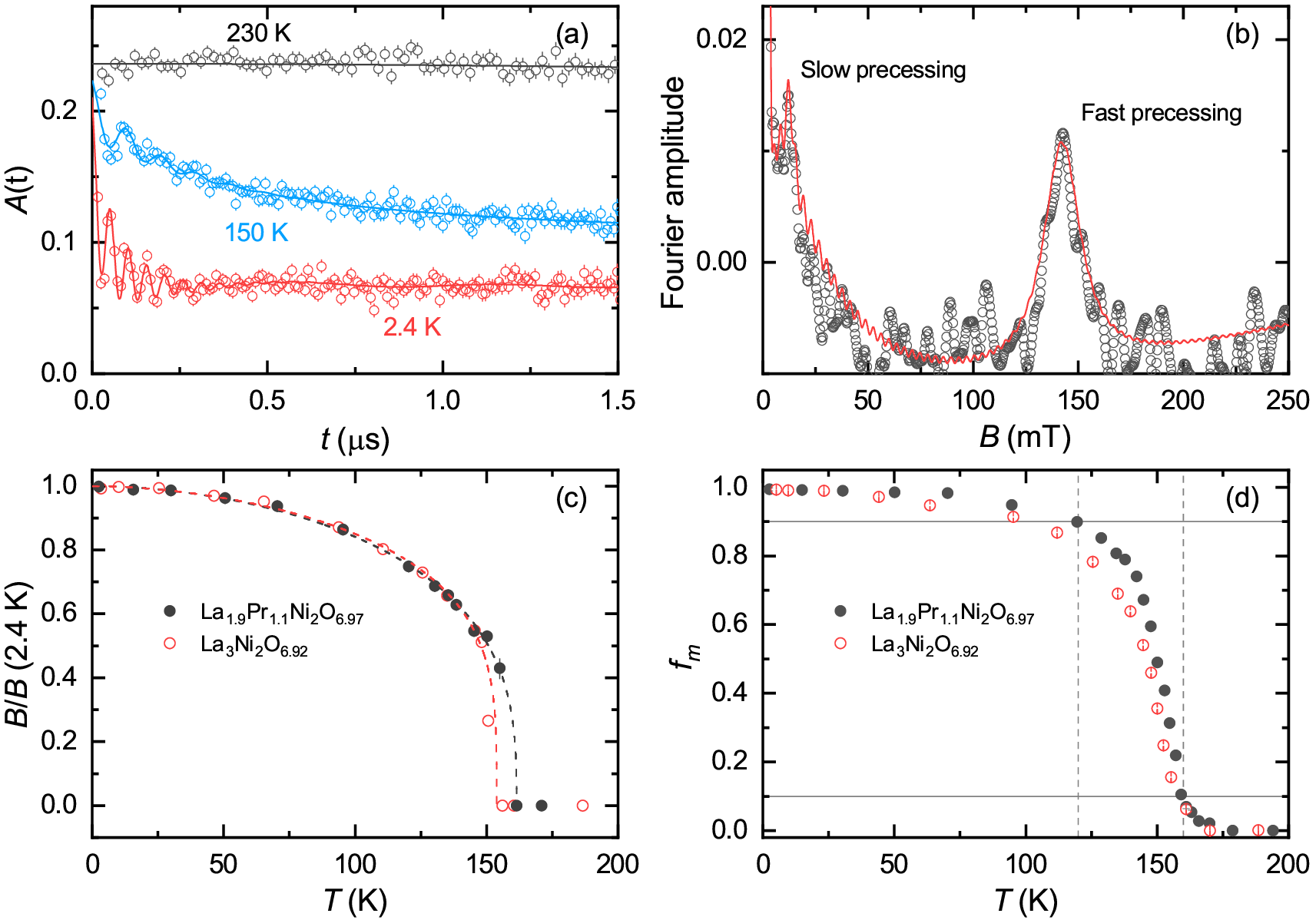}
		\caption{\label{fig:Fig3}(a) Temperature dependence of ZF-\musr\ spectra of \LPNO\ to 1.5 $\rm{\mu s}$. The solid curves are the fits with Eq.~(\ref{eq:2}). The 2.4 K data is fitted with two precessing components. (b) The Fourier transform diagram of ZF-\musr\ data measured at 2.4 K. The red line exhibit the fitting with two precessing components.  (c) Temperature dependence of magnetic order parameter $B_{\rm{int}}$ deduced from the fitting with Eq.(\ref{eq:2}). The black dashed line exhibits the fitting with Eq.~(\ref{eq:4}). (d) Magnetic volume fraction $f_m$ extracted from wTF-\musr\ experiments. The black lines mark the temperature where $f_m$ reaches 10\% and 90\%. The result of \LNO\ are cited from ref~\cite{Chen2024}.}
	\end{center}
\end{figure}

\textit{Muon spin relaxation.}--- Representative muon spin relaxation spectra of \LPNO\ are plotted in Fig.~\ref{fig:Fig3}(a). The muon spin relaxation above 185 K is primarily due to the Gaussian-distributed static nuclear dipolar moments, which can be described with a single Gaussian Kubo-Toyabe (KT) function~\cite{Blundell1999,Hayano1979}. An exponential term emerges below 170 K without detectable oscillation signals. An oscillation term emerges below 160 K, with its volume fraction increasing upon cooling. The Fourier transform diagram of 2.4 K data shown in Fig.~\ref{fig:Fig3}(a) is plotted in Fig.~\ref{fig:Fig3}(b). The frequency on the x-axis is divided by the muon gyromagnetic ratio ($\gamma_{\rm{\mu}}=851.616$ MHz/T) to be converted into magnetic field units. The Fourier spectrum at 2.4 K is manifested as two characteristic fields: $B_{\rm{slow}}\approx10$ mT and $B_{\rm{fast}}\approx140$ mT. The ZF-\musr\ spectrum is therefore described with the following formula:
\begin{multline}
\label{eq:2}
A_{\rm{ZF}}/A_0=(1-f_m)G_{\rm{KT}}(\sigma_{\rm{ZF}},t)\\
	 	+f_m[(1-f_L)P_{\perp}(t)+f_Le^{{-\lambda_L}t}],
\end{multline}
where
\begin{eqnarray}
\label{eq:3}
P_{\perp}(t)=\sum_{i=1}^{n}f_i\cos(\gamma_{\rm{\mu}}B_{\rm{int,i}}t)e^{-\lambda_it}.
\end{eqnarray}
The first term of Eq.~(\ref{eq:2}) describes the signal of paramagnetic phase and the second term describes the signal of magnetic phase. $G_{\rm{KT}}$ is the Gaussian Kubo-Toyabe function and $f_m$ is the magnetic volume fraction. $\sigma_{\rm{ZF}}=0.08$ \textmu s$^{-1}$ is fixed during the fitting. Temperature dependence of $\lambda$ and $f_i$ are plotted in supplementary material. $f_L=0.299$ is determined by the long term spectra of both ZF and wTF-\musr\ spectra. $f_L$ is slightly smaller than the expected value 1/3 in polycrystalline, which could be attributed to the spatial preferential orientation induced by the pelletization process~\cite{Sugiyama2009}. Fitting with Eq.~(\ref{eq:2}) on the data of 2.4 K is plotted in Fig.~\ref{fig:Fig3}(a). However, the volume fraction of the slow precessing term is small ($\sim$0.005) and is hard to be resolved above 95 K and we cannot determine $B_{\rm{slow}}(T)$ within an acceptable range of error (see supplementary material)~\cite{Supmat}. The spectra of the whole temperature range are finally fitted with one cosine component and one fast exponential decay term in Eq.~(\ref{eq:3}). The oscillation can be fitted with single cosine function suggests a commensurate magnetic order. Temperature dependence of normalized $B_{\rm{int}}(T)$ extracted by Eq.~(\ref{eq:2}) is plotted in Fig.~\ref{fig:Fig3}(c). $B_{\rm{int}}(T)$ is further fitted with the following phenomenological function:
\begin{eqnarray}
\label{eq:4}
B_{\rm{int}}(T)=B_{\rm{int}}(0)[1-(T/T_N)^{\alpha}]^{\beta}.
\end{eqnarray}
The fitting results in exponent $\alpha=1.9(2), \beta=0.33(2)$, $T_N=161.4(5)$ K, $B(0)=145(1)$ mT. The $\beta$-exponent obtained from fitting is close to the value of 0.326 expected in 3D Ising system~\cite{Talapov1996}. The result of La$_3$Ni$_2$O$_{6.92}$ is also plotted in Fig.~\ref{fig:Fig3}(c) for comparison. Considering the positive chemical pressure introduce by Pr-substitution, a higher $T_N=161$ K of \LPNO\ revealed by ZF-\musr\ is consistent with the result of pressurized \musr~\cite{Zurab}, indicating a split of SDW transition under pressure. Overall, the magnetization of \LPNO\ is quite different from La$_3$Ni$_2$O$_{6.92}$, which will be discussed later.

The magnetic volume fraction is extracted by wTF-\musr\ experiment. In the wTF-\musr\ experiment, an external field  perpendicular to the initial muon spin polarization direction is applied. The external field is relatively weak and will not change the internal field distribution of the magnetic phase. The muons stopping in the magnetic phase experience a broader field distribution and the muon spins depolarize rapidly. On the other hand, the muons stopping in the paramagnetic phase will precess along the external field with frequency $\omega=\gamma_{\rm{\mu}}B_{\rm{ext}}$. The wTF-\musr\ time spectra are shown in supplementary material. The spectra are fitted with: 	 
\begin{eqnarray}
\label{eq:5}
A(t)=A_{\rm{PM}}(T)e^{-\frac{\sigma^2t^2}{2}}\cos{(\gamma_{\rm{\mu}}B_{\rm{ext}}t)}+A_M(T)e^{-\lambda t}.
\end{eqnarray}
The first term describes the muons stopping in the paramagnetic phase, The second term describes the long time \textquoteleft1/3\textquoteright\ component of the magnetic phase. The fitting is limited to the data after 0.5 \textmu s to avoid the impact of the fast relaxation component of the magnetic phase. The magnetic volume fraction can be obtained by:
\begin{eqnarray}
\label{eq:6}
f_m(T)=1-A_{\rm{PM}}(T)/A_{\rm{PM}}(300 \rm{K}).
\end{eqnarray}
The obtained magnetic volume fraction is plotted in Fig.~\ref{fig:Fig3}(d), together with the result of La$_3$Ni$_2$O$_{6.92}$. The magnetic transition width of \LPNO\ is narrower, indicating a purer phase.

The \musr\ experiment results of \LNOO\ are summarized in Fig.~\ref{fig:Fig4}. wTF-\musr\ spectra are shown in supplementary materials. There is no detectable oscillation component in ZF-\musr\ spectrum. Instead, an initial  asymmetry loss is observed below 300 K. Meanwhile, ZF and wTF-\musr\ spectra at 300 K split before and after applying the field, indicating the presence of ferromagnetism at 300 K, which is also confirmed by magnetization measurement~\cite{Supmat}. The magnetic volume fraction obtained by Eq.~(\ref{eq:5}) and Eq.~(\ref{eq:6}) is plotted in Fig.~\ref{fig:Fig4}(b). $f_m(T)$ exhibits a two-step transition: one gradually reaches 50\% of total amplitude from 300 K to 30 K, and a second sharper transition below 30 K, finally reaches the value of 95\%.  To clarify the dynamic characteristic and magnitude of magnetism, longitudinal field were applied at 130 K and 2.5 K. A longitudinal field ten times larger than the static internal field can fully decouple the muon spins from the internal field, which is manifested as the suppression of muon spin relaxation in the spectrum. On the other hand, a dynamic internal field is much more robust to the longitudinal field~\cite{Hayano1979}. The LF-\musr\ spectra at 130 K and 2.5 K are shown in Fig.~\ref{fig:Fig4}(c) \& (d), respectively. At 130 K, the muon spins are totally decoupled from the local field under 10 mT longitudinal field, indicating a static internal field of 1 mT order or even smaller. This rules out the magnetism originating from 214 or 327 phases, whose internal fields sensed by muons are on the order of 100 mT~\cite{Chow1996,Jest1999,Chen2024}. On the other hand, it is unlikely that the signal, which accounts for 20\%\ of the total amplitude, entirely originates from ferromagnetic impurities. At 2.5 K, a 10 mT longitudinal field has almost no effect on the muon spin relaxation behavior. The loss of initial asymmetry is gradually suppressed by the longitudinal field on the order of 100 mT, indicating a static internal field of the same order of magnitude. 
	
Based on the above information, the ZF-\musr\ spectra above 30 K are fitted with:
\begin{eqnarray}
\label{eq:7}
A(t)=A_0(T)G_{\rm{KT}}(\sigma,t)e^{-\lambda t}.
\end{eqnarray}
The KT term describes the muon spin relaxation due to the nuclear dipolar moments and the exponential decay term describes the additional depolarization rate due to the randomly distributed magnetic moment discussed later. $\sigma=0.07$ \textmu s$^{-1}$ is fixed during the fitting. The ZF-\musr\ spectra below 30 K are fitted by:
\begin{eqnarray}
\label{eq:8}
A(t)=A_{\rm{fast}}e^{-\lambda_{\rm{fast}}t}+A_{\rm{tail}}e^{-\lambda_{\rm{tail}}t}.
\end{eqnarray}
The first term describes the early time fast relaxation and the second term describes the \textquoteleft1/3\textquoteright\ tail of the magnetic signal. The fitting is shown in Fig.~\ref{fig:Fig4}(a). The obtained ZF-\musr\ fitting parameters are summarized in supplementary material~\cite{Supmat}.

\begin{figure}[h]
\begin{center}
\includegraphics[width=0.5\textwidth]{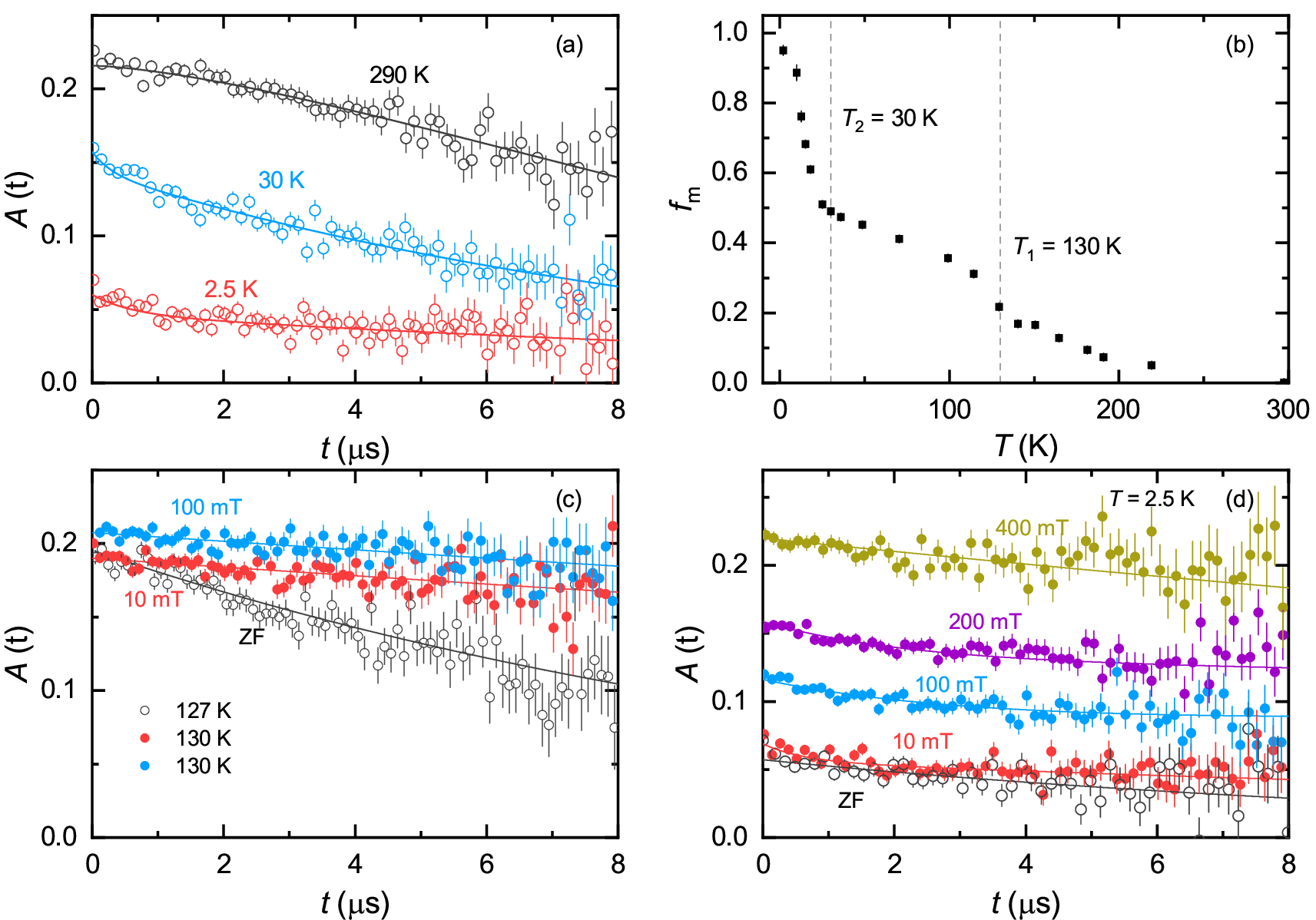}
\caption{\label{fig:Fig4}(a) Temperature dependence of ZF-\musr\ spectra of \LNOO\ to 1.5 $\rm{\mu s}$. The solid curves are the fits with Eq.~(\ref{eq:7}). (b) Temperature dependence of magnetic volume fraction $f_m(T)$ obtained by Eq.~(\ref{eq:8}).  (c) \& (d) LF-\musr\ spectrum measured at 130 K and 2.5 K, respectively. The solid lines are guides for eyes.}
\end{center}
\end{figure}

\textit{Discussion.}---The microscopic magnetization of \LPNO\ revealed by ZF-\musr\ is quite different from La$_3$Ni$_2$O$_{6.92}$. First, no second high-field precession frequency is observed in \LPNO. This can be explained by a higher phase purity and a more uniform oxygen distribution. Second, the $\beta$-exponent changes from 0.26 to 0.33. Though substitution of Pr for La introduces positive chemical pressure, similar changes were not observed in the \musr\ experiment under pressure~\cite{Zurab}. Therefore, such difference cannot be simply attributed to the reduction of interlayer distance. Given that TGA results show that \LPNO\ is nearly free from oxygen deficiency, one explanation is that a deficient-free structure can significantly enhance the interlayer coupling, thus causing the magnetization behavior to transition from two-dimensional to three-dimensional Ising case. Third, the amplitude of 10 mT signal is greatly reduced in \LPNO, which challenges the previous hypothesis of an intrinsic spin-charge structure. The low frequency may originate from the randomly-distributed oxygen vacancies, where the magnetic moments of the nearest Ni atoms diminish to nearly zero, forming effectively charge sites~\cite{Cao2024}. 
	
The two-step magnetic transition in \LNOO\ can also be understood in terms of oxygen vacancies. As revealed by LF-\musr\ experiments, the magnitude of the magnetism below and above 30 K differ by at least one order of magnitude, indicating different origins. The magnetism over 30 K can originate from the magnetic moment localized near the oxygen vacancies~\cite{Liu2024}. These moments can further form short-range spatial correlations, leading to the loss of initial asymmetry. The second transition below 30 K originates from the intrinsic magnetism of the 327 phase. $\delta=0.37$ indicates $\sim40\%$ of the inner apical vacancies. The long-range order is therefore disrupted by the oxygen vacancies and lead to a short-range order, leading to a further loss of initial asymmetry $A_0$~\cite{Cao2024}.
	
\textit{Conclusion.}---In conclusion, the microscopic magnetism of \LPNO\ and \LNOO\ has been systematically studied using \musr\ technique. Bulk commensurate long-range magnetic order is observed below 161 K in \LPNO, which has the perfect lattice structure and is nearly free from oxygen vacancies. The $\beta$-exponent is close to the 3D Ising model, indicating an enhanced interlayer coupling. On the other hand, two types of short-range magnetic orders are observed in \LNOO. The magnetism over 30 K is believed to be related to the magnetic moments localized near the oxygen vacancies, and the order below 30 K manifests the intrinsic magnetic order of La$_3$Ni$_2$O$_{7-\delta}$ phase. The long-range order is completely disrupted by the oxygen vacancies, resulting in the experimental observation of short-range characteristics. Our \musr\ results provide a comprehensive understanding of the variation of magnetism with oxygen vacancies and structural perfection in bilayer RP phase nickelate. The overall evolution of ambient-pressure magnetism exhibits consistent characteristics with the sample-dependent superconducting properties under high pressure. This further proves the importance of magnetic interactions to superconductivity in bilayer nickelate.

We are grateful to B. Hitt and D. Arsenau of the TRIUMF CMMS for assistance during the experiments. This research was funded by the National Key Research and Development Program of China, No. 2022YFA1402203, the National Natural Science Foundations of China, No. 12174065, Innovation Program for Quantum Science and Technology (Grant No. 2024ZD0300104), and the Shanghai Municipal Science and Technology Major Project (Grant No. 2019SHZDZX01). Y. F. Guo acknowledges the National Key R\&D Program of China (Grant No. 2023YFA1406100) and the Double First-Class Initiative Fund of ShanghaiTech University.

\end{document}